\documentclass{iaus}
\usepackage{graphicx}
\usepackage{amssymb}

\newcommand\etal{\mbox{\textit{et al.}}}

\title[Exponential bulges and antitruncated disks] 
{Exponential bulges and antitruncated disks in lenticular galaxies}

\author[Silchenko]   
{Olga K. Sil'chenko$^1$}%

\affiliation{$^1$Sternberg Astronomical Institute, Moscow 119991, Russia
\break email: olga@sai.msu.su \\[\affilskip]
}

\pubyear{2008}
\volume{254}  
\pagerange{xxx--yyy}
\date{?? and in revised form ??}
\jname{The Galaxy Disk in Cosmological Context}
\editors{J. Andersen et al., eds.}
\begin{document}

\maketitle

\begin{abstract}
The presence of exponential bulges and anti-truncated disks has been noticed in
many lenticular galaxies. In fact, it could be expected because the very
formation of S0 galaxies includes various processes of secular evolution.
We discuss how to distinguish between a pseudobulge and an anti-truncated
disk, and also what particular mechanisms may be responsible for the
formation of anti-truncated disks. Some bright examples of lenticular
galaxies with the multi-tiers exponential stellar structures are presented,
among them -- two central group giant S0s seen face-on and perfectly
axisymmetric. 
\end{abstract}

\firstsection 
\section{Introduction}

Bulges have been traditionally thought to have de Vaucouleurs' brightness
profiles just as elliptical galaxies (e.g. \cite[Freeman 1970]{freeman}). 
However John Kormendy (\cite[Kormendy 1982a]{korm82a}, 
\cite[Kormendy 1982b]{korm82b}, \cite[Kormendy 1993]{korm93})
has proved that there exists a type of bulges named `pseudobulges' which
resemble disks from the dynamical point of view: they are rather cold and
demonstrate fast rotation. Despite their dynamical properties, they are bulges
from the geometrical point of view: they are `fat' and have rather large
scaleheights. Pseudobulges are thought to be products of secular evolution;
but if it is so, then dynamical simulations 
(e.g. \cite[Pfenniger \& Friedli 1991]{pf_fri}) predict that
they must have exponential brightness profiles. And indeed, small bulges
of late-type spirals which could be easily made by secular evolution, 
practically all have exponential brightness profiles 
(e.g. \cite[Andredakis et al. 1995]{phot_apb}, \cite[Graham 2001]{graham}).
But it is not only small bulges in late-type spirals 
that have exponential brightness profiles. Lenticular galaxies 
whose bulges are not usually small at all demonstrate typically 
the mean Sersic coefficient lesser than Sa galaxies, and often their
$n  <2$: the Sersic parameter
$n$ peaks at the Sa--Sb morphological type and falls further toward S0s 
(\cite[Graham 2001]{graham}, \cite[Laurikainen et al. 2005]{lauri05},
\cite[Laurikainen et al. 2007]{lauri07}). In fact, it is quite natural 
because the very event of a S0 galaxy
transformation from a spiral must include various processes 
of secular evolution resulting in matter re-distribution over 
the radius and bulge reshaping.

But galactic disks outside the bulges can also consist of several exponential
segments. Now it becomes clear that so called anti-truncated disks 
consisting of two exponential segments with the outer scalelengths larger
than the inner ones, may be the dominant type of galactic disks among 
some types of galaxies (\cite[Erwin et al. 2008a]{erwinstat}).
We reported finding a few large nearby spiral galaxies with such
two-tiers disks during the last 10 years.
A significant population of such disks has been found since the brightness
profiles begin to reach 27th and 28th magnitudes from one square arcsecond
(\cite[Pohlen \& Trujillo 2006]{pohtru}).
Does it mean that the outer segments of anti-truncated disks represent always
low surface-brightness (LSB) disks? The statistics by Erwin et al. 
(\cite[Erwin et al. 2008b]{erwinbar}) for barred galaxies says so. 
Among our findings in unbarred spiral galaxies, 
only one galaxy, NGC~5533, has a LSB outer disk 
(\cite[Sil'chenko et al. 1998]{n5533}). 
Other galaxies 
(NGC~7217, \cite[Sil'chenko \& Afanasiev 2000]{n7217}; 
NGC~615, \cite[Sil'chenko et al. 2001]{n615}; 
NGC~4138, \cite[Afanasiev \& Sil'chenko 2002]{we2002}; 
NGC~7742, \cite[Sil'chenko \& Moiseev 2006]{n7742}) have on the
contrary quite normal outer disks, if to compare with the Freeman's 
(\cite[Freemen 1970]{freeman})
reference value of the central $B$ surface brightness of $21.7$; and
simultaneously they have compact bright inner disks. Can we distinguish
the compact inner disk from a pseudobulge? Yes, by using multi-variant
approach. For example, in NGC~7742 all spirals and current star formation
are confined to the inner exponential disk, and the outer one is smooth and 
featureless (\cite[Sil'chenko \& Moiseev 2006]{n7742}), 
so we can conclude that the inner
exponential stellar component is dynamically cold and cannot be a bulge. 
But the key property is a visible geometry
of the stellar component with the exponential profile. To be a disk, it
must be thin. For disks inclined to the line of sight, a good check is
isophotal analysis. The outer disk is always assumed to be thin: then its 
isophote axis ratio characterizes the cosine of the inclination. If the
inner component has the same visible axis ratio as the outer one and if
two disks are coplanar, the inner structure cannot be a spheroid, it must 
be a disk. For face-on galaxies, this approach does not work: their
isophotes are always round independently of the scaleheights. But for
face-on galaxies we can use kinematical data and estimate their thickness
by measuring vertical stellar velocity dispersions.

We have now started a program of studying systematically multi-tiers 
(anti-truncated) exponential structures in early-type, presumably lenticular, 
galaxies. The study will include photometric as well as spectral observations. 
Some first results are presented in this talk.

\section{Observations}

The photometric observations the results of which we discuss here have been
made with the focal reducer SCORPIO of the Russian 6m telescope
(\cite[Afanasiev \& Moiseev 2005]{scorpio}) in the direct-image mode. 
The CCD detector EEV 42-40 with the size of $2048 \times 2048$ 
has been used in binned mode of $2\times 2$. The field of view 
was about 6 arcminutes, the scale was 0.35 arcsec per binned pixel. 
The photometric observations have been undertaken
on August 21, 2007, under the seeing quality of about $2^{\prime \prime}$.
We have exposed 5 fields in the NGC~80 group and 6 fields in the
NGC~524 group, through two filters, $B$ and $V$. The exposure times
were selected in accordance with the surface brightness of the
targets observed; for example, we exposed NGC~524 itself during
60 sec in the $B$-filter and during 30 sec in the $V$-filter.
As a flat field, we used the exposures of the twilight sky.
The calibration onto the standard Johnson $BV$ system has been made
by using multi-aperture photoelectric data collected by HYPERLEDA
for NGC~80 and NGC~524.

\section{Photometric structure of the central group S0 galaxies}

Two central group galaxies under consideration are typical
giant lenticular galaxies, with the blue absolute magnitudes of
about $-21.6 - -21.7$ (HYPERLEDA). Both are very red, $(B-V)_e=1.07$, 
and are seen face-on, $b/a>0.9$.

\begin{figure*}
\begin{centering}
\includegraphics[height=4.3cm]{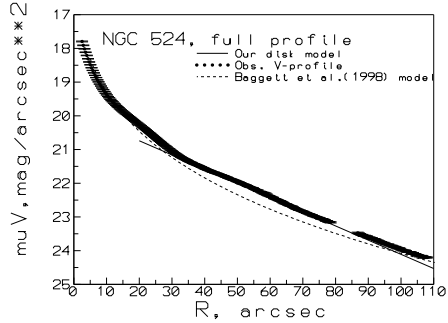}
  \caption{The V-band azimuthally-averaged brightness profile of NGC 524 
obtained by us with the reducer SCORPIO at the 6m telescope; the model
decomposition proposed by Baggett et al. (1998) is overlaid. One can see
that the high-accuracy data do not agree with a single de Vaucouleurs'
bulge model; instead at least two exponential components are needed.}
\end{centering}
\label{n524vprof}
\end{figure*}

\begin{figure*}
\includegraphics[width=4.3cm]{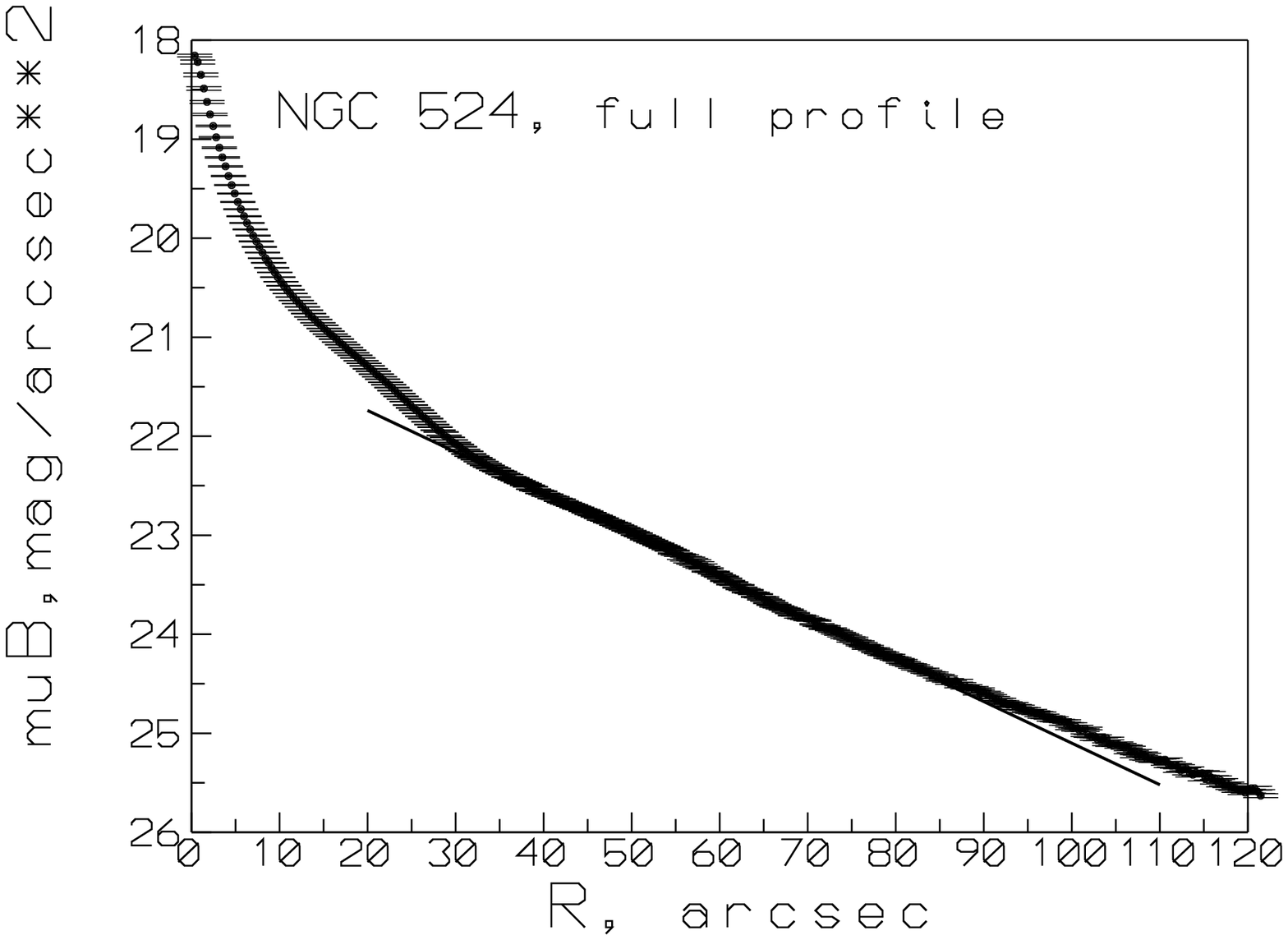}
\includegraphics[width=4.3cm]{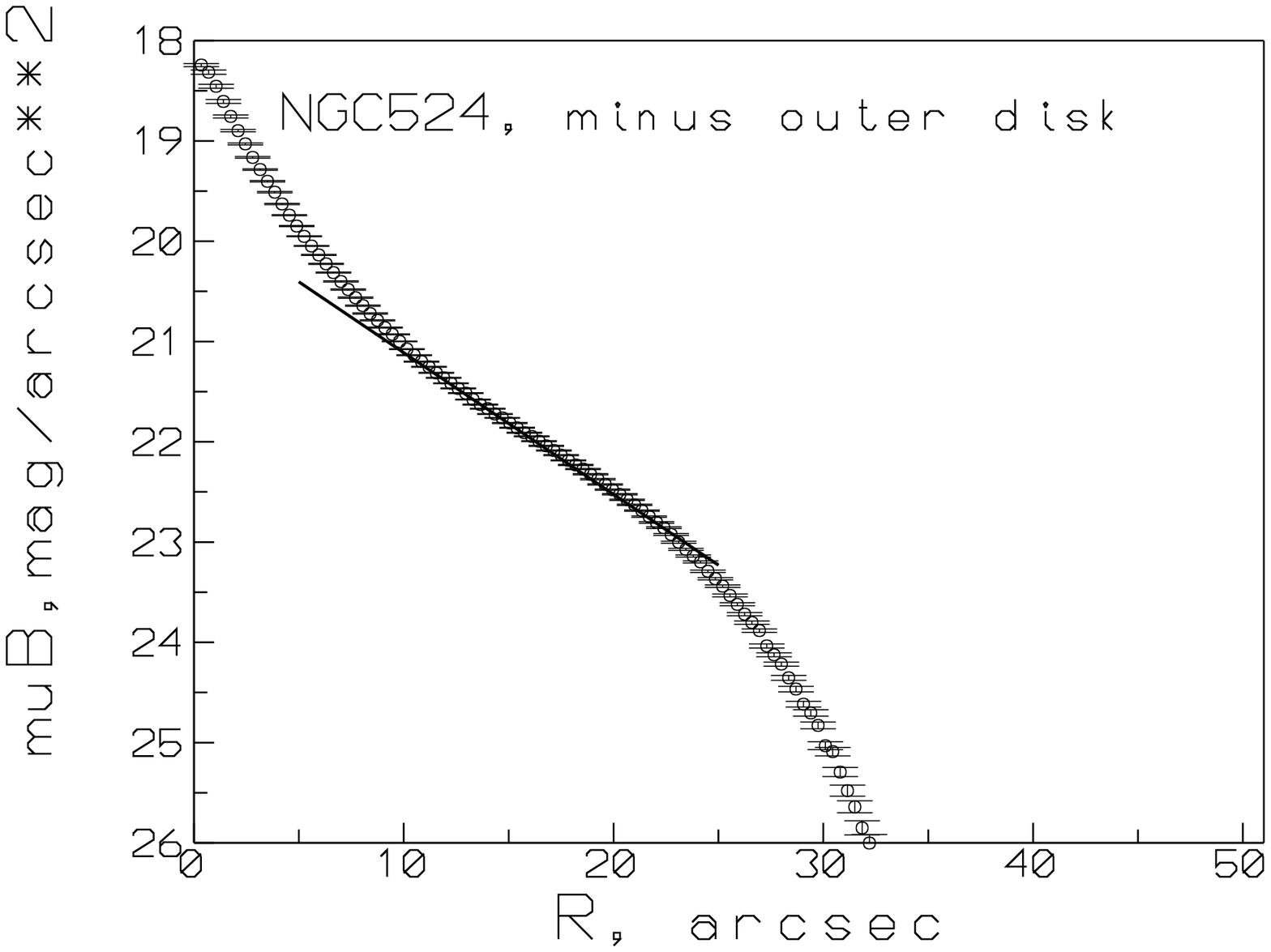}
\includegraphics[width=4.3cm]{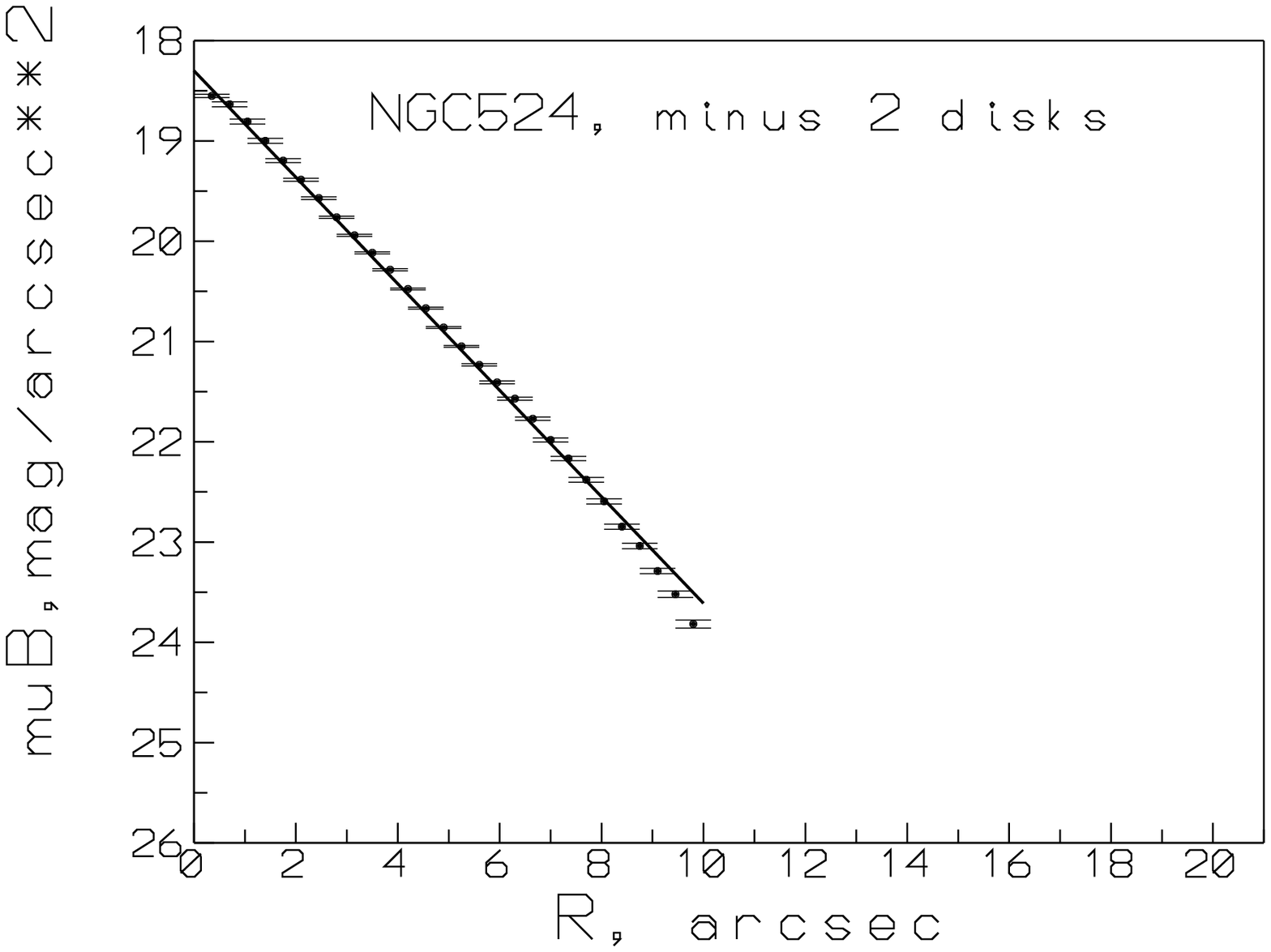}
  \caption{The B-band azimuthally-averaged brightness profile of NGC 524 
obtained by us with the reducer SCORPIO at the 6m telescope can be 
decomposed step-by-step into three exponential components.}
\label{n524bprof}
\end{figure*}

We have calculated azimuthally averaged surface brightness profiles for
NGC~524 in two filters, $B$ and $V$. The data are rather precise, and we trace
the profiles up to $R=80^{\prime \prime}$, or about 10 kpc from the center,
with the accuracy better than 0.01 mag. At larger radii the accuracy is worse
due to bright stars projected onto the galaxy. When compare our $V$-profile
with the model decomposition proposed by Baggett et al. 
(\cite[Baggett et al. 1998]{baggett}) (Fig.~\ref{n524vprof}), 
namely, with a single de Vaucouleurs' bulge approximation, one can see that 
their model is inappropriate. Instead we see at least two exponential components: 
the `outer' one in the radius range of $R=35^{\prime \prime} -90^{\prime \prime}$
and the `inner' one in the radius range of $R=10^{\prime \prime} -25^{\prime \prime}$.
We fit the outer part of the surface brightness profile by an exponential law,
construct a 2D model image of this disk and subtract it from the full image observed.
For the residual image, we calculate again the azimuthally averaged surface
brightness profile, fit its outer part by an exponential law, construct the 2D image
of the inner disk and subtract it from the first-step residual image. Interestingly
the brightness profile of the second-step residual image which is safely
traced up to $10^{\prime \prime}$, is also exponential (Fig.~\ref{n524bprof})!
Figure~\ref{n524bprof} presents all steps of our decomposition procedure, and in
the Table 1 we give the parameters of the exponential stellar substructures
obtained by this procedure.

\begin{table*}
\caption[ ] {Parameters of the photometric components of NGC 524
approximated by an exponential law}
\begin{flushleft}
\begin{tabular}{|l|cccc|}
\hline
Component & Radius range of the fit &
$\mu_0$, mag/$\square ^{\prime \prime}$
& $h^{\prime \prime}$ & $h$, kpc \\
\hline
\multicolumn{5}{|c|}{$B$-filter}\\
Outer disk & $34^{\prime \prime} - 90^{\prime \prime}$
& $20.906 \pm 0.007$ & $26.07 \pm 0.06$ & 3.02 \\
Inner disk & $10^{\prime \prime} - 23^{\prime \prime}$
& $19.67 \pm 0.01$ & $7.67 \pm 0.05$ & 0.89 \\
Bulge & $1^{\prime \prime} - 8^{\prime \prime}$
& $18.25 \pm 0.01$ & $2.04 \pm 0.01$ & 0.24 \\
\hline
\end{tabular}
\end{flushleft}
\end{table*}

\begin{figure*}
\begin{centering}
\includegraphics[width=4.3cm]{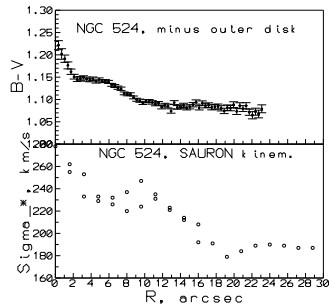}
  \caption{The comparison of the azimuthally averaged profiles
of the $B-V$ colour calculated from the residual image of NGC 524
after the subtraction of the outer disk and of the stellar
velocity dispersion demonstrating a break at the edge of a
dynamically hot bulge.}
\end{centering}
\label{n524cvet}
\end{figure*}

Figure~\ref{n524cvet} presents the $B-V$ colour profile of the 
first-step residual image
(the inner disk). It reveals a net colour difference between the inner
disk, $R>10^{\prime \prime}$, and the `bulge', $R<10^{\prime \prime}$.
This colour difference, $\Delta (B-V) =0.07$, may be resulted from a 
metallicity difference of 0.2 dex under the old stellar age $T=12$ Gyr,
or on the contrary from an age difference of 8 Gyr under the metallicity
of [Z/H]$=+0.1$ 
(\cite[Worthey 1994]{worthey}, and also his WEBsite, Dial-a-Galaxy option),
the inner disk being younger or/and more metal-poor. 
We have compared the colour profile
of the residual image with the stellar velocity dispersion profile
similarly averaged in the rings over the map obtained from the SAURON
data (\cite[Emsellem et al. 2004]{sau3}, 
see also the WEBsite of the SAURON project) (Fig.~\ref{n524cvet}).
The profiles of the colour and of the stellar velocity dispersion are
qualitatively similar! Certainly, we see a transition from the
(exponential) bulge to the inner disk at $R\approx 10^{\prime \prime}$.
Preliminary estimates of the scaleheight of the inner disk in NGC~524,
by treating the measured line-of-sight stellar velocity dispersion as
a vertical one, have given a value of about 1.5 kpc; it is a typical
scaleheight of a thick stellar disk which can be expected in a lenticular
galaxy.

The very similar surface brightness profile consisting of two exponential
disks and of a very compact bulge has been observed by us earlier
in the giant lenticular galaxy NGC~80, which is also settled at the
center of a rich X-ray group (\cite[Sil'chenko et al. 2003]{n80}).
With the new observations, we confirm certainly this
multi-tiers structure. In both galaxies the outer stellar disks are quite
normal as concerning their scalelengths or their central surface brightnesses,
and the inner disks are compact and bright. Both galaxies are seen face-on 
and are certainly unbarred; the low ellipticity of their isophotes over 
the full radial extension proves that the galaxies are strictly axisymmetric. 
If the exponential inner stellar structures have been formed by secular
evolution, where are signatures of the main `driver' of secular evolution,
of a bar?

\section{What can be the mechanisms to form anti-truncated disks in
lenticular galaxies?}

It seems clear that anti-truncated disks are to be a result of matter
re-distribution along the radius of a disk galaxy, and the very event of 
re-distribution must be rather fast and discrete. Several candidate 
mechanisms can be proposed. Younger et al. 
(\cite[Younger et al. 2007]{sim_anti}) simulates a minor 
merger, and they obtain an anti-truncated stellar disk in the merger remnant,
mainly due to stellar diffusion from the inner part of the initial spiral
galaxy into an outer region. In such model the outer part of the
multi-tiers stellar disk in the merger remnant seems to be a low
surface brightness disk. Some years ago I proposed another mechanism
where a transient interaction, due to, say, by-pass of a rather massive
galaxy, provokes intense gas inflow and results in gas concentration
in the very inner part of a galaxy with the subsequent star formation
burst. In the frame of this model, the inner disk must be more bright
and compact than the usual large-scale stellar disks of spiral galaxies are.
We may expect that an observational statistics of the parameters of
inner and outer exponential disks in the anti-truncated galaxies would help
to select a model.

In the group NGC~80, besides the central galaxy, we have analysed brightness
profiles in more 10 lenticular galaxies, with the absolute blue magnitudes
from --17 to --20. Among those, 7 S0s have appeared to possess two-tiers
exponential disks. And among 7 S0 galaxies with the two-tiers disks, three
have the inner compact disks and the outer normal disks, and four have
the inner normal disks and the outer LSB ones. Together with NGC~80 itself
and with NGC~524, with their inner compact bright disks and extended normal
ones, we obtain half-to-half preliminary statistics which implies that 
the origin of the multi-tiers exponential disks may be different 
in different galaxies.

\section{Conclusions}\label{sec:concl}

If we assume that multi-tiers exponential profiles are formed by secular evolution
of galactic disks, the best place to search for them would be lenticular galaxies.
Lenticulars galaxies had to reform secularly their stellar disks during their 
transformation from S to S0; hence S0s must be the hosts of both multi-tiers disks 
and pseudobulges. To select a particular mechanism of forming anti-truncated disks,
we must know better the statistics of their properties which are not still 
collected over representative samples. Among the possible alternatives
of the origin of multi-tiers exponential profiles there
are minor merger versus tides (but the same alternative exists for the origin of 
S0s beyond clusters!). To choose, we need to know a balance between
compact$+$normal and normal$+$LSB stellar disk combinations.

Secular redistribution of stars and other matter in disks is thought
to be made by bars: bars are usually generated by any interaction
and even without interactions -- by intrinsic instabilities.
But almost all our galaxies with the multi-tiers disks are unbarred;
the face-on S0s NGC~524 and NGC~80 are perfectly axisymmetric.
And the samples of Erwin et al. 
(\cite[Erwin et al. 2008a]{erwinstat}, \cite[Erwin et al. 2008b]{erwinbar}) 
imply the same conclusion: among barred galaxies, the anti-truncated disks 
contribute one third of all, among unbarred -- more than 50\%. 
This inconsistency is a complete puzzle yet.

\begin{acknowledgments}
The 6m telescope is operated under the financial support of 
Science and Education Ministry of Russia (registration number 01-43). 
During our data analysis
we used the Lyon-Meudon Extragalactic Database (HYPERLEDA) supplied by the
LEDA team at the CRAL-Observatoire de Lyon (France) and the NASA/IPAC
Extragalactic Database (NED) operated by the Jet Propulsion
Laboratory, California Institute of Technology under contract with
the National Aeronautics and Space Administration. 
The study of multi-tiers galactic disks is supported by the grant of 
the Russian Foundation for Basic Researches (RFBR), no. 07-02-00229a.
My attendance at the IAU Symposium no. 254 is due to the IAU grant.
\end{acknowledgments}

\end{document}